\documentclass[10pt, a4paper, twocolumn, twoside]{IEEEtran}

\usepackage{cite}
\usepackage{enumerate}
\usepackage{graphicx}
\usepackage[cmex10]{amsmath} 
\usepackage{amssymb}
\usepackage{xspace}
\usepackage{subfigure}
\usepackage[lined,linesnumbered,ruled,commentsnumbered]{algorithm2e} 
\usepackage{colonequals}
\usepackage[mathscr]{euscript}
\usepackage{fixltx2e}    
\usepackage{amsthm}
\usepackage{mathrsfs}

\usepackage{xspace}

\usepackage{epsfig}
\usepackage{epstopdf}

\usepackage{tikz}
\usepackage{pgfplots}

\usetikzlibrary{arrows,shapes,backgrounds,plotmarks,positioning}

\pgfplotsset{compat=newest}                         
\pgfplotsset{plot coordinates/math parser=false}
\newlength\figureheight
\newlength\figurewidth

\newtheorem{example}{Example}

\newcommand{\RZ}[1]{\mathsf{Z}_{#1}}

\newcommand{\RW}[1]{\mathsf{W}_{#1}}

\newcommand{\flow}{\varphi}
\newcommand{\boundary}{\partial\varphi}
\newcommand{\Boundary}{\boldsymbol{\partial\varphi}}

\newcommand{\dep}{\text{dep}}

\newcommand{\Cut}{\kappa}

\newcommand{\Card}[1]{|#1|}
\newcommand{\Set}[1]{\{#1\}}

\newcommand{\Real}{\mathbb{R}}
\newcommand{\RealP}{\mathbb{R}_{+}}    
\newcommand{\RealPP}{\mathbb{R}_{++}}    
\newcommand{\Z}{\mathbb{Z}}            

\newcommand{\SFM}{\text{SFM}}

 {\begin{list}{}%
         {\setlength{\leftmargin}{#1}}%
         \item[]%
 }
 {\end{list}}

\begin{document}

\title{Distributed Data Compression in Sensor Clusters: A Maximum Independent Flow Approach}

\author{Ni~Ding\IEEEauthorrefmark{1}, Parastoo~Sadeghi\IEEEauthorrefmark{2}, David~Smith\IEEEauthorrefmark{1} and Thierry~Rakotoarivelo\IEEEauthorrefmark{1}

\thanks{\IEEEauthorblockA{\IEEEauthorrefmark{1}Ni Ding, David Smith and Thierry Rakotoarivelo are with Data61 (email: $\{$ni.ding, david.smith, thierry.rakotoarivelo$\}$@data61.csiro.au).}}
\thanks{\IEEEauthorblockA{\IEEEauthorrefmark2}Parastoo Sadeghi is with the Research School of Engineering, College of Engineering and Computer Science, the Australian National University (email: $\{$parastoo.sadeghi$\}$@anu.edu.au). }
}


\maketitle

\begin{abstract}
Let a cluster (network) of sensors be connected by the communication links, each link having a capacity upper bound. Each sensor observes a discrete random variable in private and one sensor serves as a cluster header or sink. Here, we formulate the problem of how to let the sensors encode their observations such that the direction of compressed data is a feasible flow towards the sink. We demonstrate that this problem can be solved by an existing maximum independent flow (MIF) algorithm in polynomial time. Further, we reveal that this algorithm in fact determines an optimal solution by recursively pushing the remaining randomness in the sources via unsaturated communication links towards the sink. We then show that the MIF algorithm can be implemented in a distributed manner. For those networks with integral communication capacities, we propose an integral MIF algorithm which completes much faster than MIF. Finally, we point out that the nature of the data compression problem in a sensor cluster is to seek the maximum independent information flow in the intersection of two submodular polyhedra, which can be further utilized to improve the MIF algorithm in the future.
\end{abstract}


\section{introduction}
\begin{figure}[tpb]
	\centering
    \scalebox{1}{\begin{tikzpicture}

\draw (0,0) circle (0.3);
\node at (0,0) {\Large $2$};

\draw (1.5,1) circle (0.3);
\node at (1.5,1) {\Large $3$};

\draw (1.5,-1) circle (0.3);
\node at (1.5,-1) {\Large $t$};

\draw (-1.5,1) circle (0.3);
\node at (-1.5,1) {\Large $1$};

\draw (-1.5,-1) circle (0.3);
\node at (-1.5,-1) {\Large $4$};

\draw [->] (-1.2,1) -- node [auto] {\scriptsize $2$} (1.2,1);
\draw [->] (1.5,0.7) -- node [auto] {\scriptsize $2$} (1.5,-0.7);
\draw [->] (-1.3,0.8) -- node [sloped,above] {\scriptsize $1$} (-0.25,0.15);
\draw [->] (-1.5,0.7) -- node [auto,left] {\scriptsize $3$} (-1.5,-0.7);
\draw [->] (-1.3,-0.8) -- node [sloped,above] {\scriptsize $1$} (-0.25,-0.15);
\draw [->] (0.25,-0.15) -- node [sloped,above] {\scriptsize $0.6$}(1.3,-0.8);
\draw [->] (1.2,-1) -- node [above] {\scriptsize $1$}(-1.2,-1);

\end{tikzpicture} }
	\caption{A digraph that represents a sensor cluster: There are five sensors, $1,\dotsc,4$ and $t$. For $i,j \in \Set{1,\dotsc4,t}$, the edge $(i,j)$ represents a communication link from node $i$ to node $j$. There is a flow upper bound $c(i,j)$ associated with each edge $(i,j)$, e.g., $c(1,2) = 1$. Node $t$ is selected as the cluster header to collect all the measurements from other sensor nodes.}
	\label{fig:Digraph}
\end{figure}
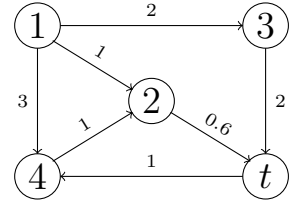

Emerging studies on wireless sensor networks and their applications pose new challenges to the data compression problem. A sensor network is usually sectioned into clusters, e.g., based on geographic location, and, in each cluster, a sensor node is selected as the cluster header to collect all sensing data from others \cite{Yick2008WSN}. The sensor nodes in a cluster are assumed to be connected by communication links so that each sensor node not only sources information (i.e., record measurements/observations) from the environment but also relays/forwards the incoming compressed data from other nodes at the same time \cite{Yu2005}. See Fig.~\ref{fig:Digraph}. It is also shown in \cite{Vuran2004} that there is a strong spatial-temporal correlation in the sensing data. Then, there is a data compression problem of how to determine the source coding rate for each sensor to encode its measurements/observations so that the compressed data can be successfully forwarded over the communication links to the cluster header.

This multiterminal source coding problem in a sensor cluster/network has been studied in \cite{Cristescu2004,Cristescu2005,Aditya2011}. The authors in \cite{Cristescu2004,Cristescu2005} proposed a two-step approach: determine the (minimum) spanning tree of the sensor network and apply the Edmond greedy algorithm \cite{Edmonds2003Convex} to determine an extreme point in the Slepian-Wolf achievable source code rate region \cite{SW1973,Cover1975} for the lossless data compression. But, this approach does not exploit all the communication resources: Since the communication links are wireless, any outgoing links of a node, not just the ones in the spanning tree, can be utilize to forward the compressed data.
On the other hand,  a combinatorial optimization problem is formulated in \cite{Aditya2011}. But, instead of utilizing the submodularity of the data compression and routing problem,\footnote{The entropy function in the data compression problem and the cut function, which determines the maximum flow in a graph, are both submodular, the minimization of which can be solved in polynomial time \cite[Sections 1.2 and 2.2]{Fujishige2005}. } the optimal solution is determined by a centralized subgradient based algorithm, a discrete optimization technique.

In this paper, we model the sensor cluster by a capacitated multiple-source-single-sink  digraph, where there is a flow upper bound applied to each communication link, e.g., Fig.~\ref{fig:Digraph}. We assume that each source/sensor node observes a component of a discrete memoryless multiple source (DMMS) in private and we consider the problem of how to let the source nodes encode their observations so that the compressed data can be directed as a feasible flow towards the sink. We show that this problem can be directly solved by a maximum independent flow (MIF) algorithm \cite{FujishigeMaxIndFlow1978} which is based on the submodular function minimization (SFM) techniques \cite[Chapter VI]{Fujishige2005} and completes in polynomial time.
We show how to implement the MIF algorithm in a distributed manner and explain that the MIF algorithm in fact determines an optimal solution by recursively pushing the remaining randomness in the sources via unsaturated communication links towards the sink. Based on this interpretation, we propose an integral MIF (IMIF) algorithm for determining an integral optimal solution when the capacities are integral and the entropy function of the DMMS is integer-valued. We show that the complexity of the IMIF algorithm is much less than the MIF algorithm. Finally, we point out that the nature of the data compression problem in a sensor cluster is to seek the maximum independent information flow in the intersection of two submodular polyhedra, the mathematical results of which can be further utilized to improve the MIF algorithm in the future.

\section{System Model}
\label{sec:system}

For a finite set $V$ with $\Card{V}>1$, let $G=(V \cup \Set{t}, E, c)$ be a digraph that is connected.\footnote{In this paper, a digraph is called connected if there is a path between any two nodes $i,j \in V$ in the underlining undirected graph.}
The node set $V \cup \Set{t}$ contains all the indices of the sensors in a cluster with sensor node $t$ being the cluster header or the sink. The edge set $E$ contains all the communication links in the cluster: There is an edge $(i,j) \in E$ if $j$ is in the communication range of $i$. The capacity function is $c \colon E \mapsto \RealPP$ and $c(i,j)$ denotes the flow upper bound on edge $(i,j)$. For node $i$, the sum inflow capacity $\sum_{j \in V \colon (j,i) \in E} c(j,i)$ indicates the processing capability of $i$, the  maximum inflow information amount (e.g., in bits) that can be processed by node $i$. For example, in the digraph $G$ in Fig.~\ref{fig:Digraph}, we have $\sum_{j \colon (j,2) \in E} c(j,2) = 2$ state sensor node $2$ can only relay/process a maximum of 2 bits incoming compressed data in addition to the randomness in its own observations.
A \textit{flow} $\flow \colon E \mapsto \RealP$ assigns each edge a nonnegative value. We say that $\flow$ is a \textit{feasible flow} in $G$ if $f(i,j) \leq c(i,j),\forall (i,j) \in E$.

For each $i\in V$, sensor $i$ observes an i.i.d.\ $n$-sequence $\RZ{i}^n$ of the discrete random variable $\RZ{i}$ in private. The observations are in general correlated so that all $\RZ{i}$s form a discrete memoryless multiple source (DMMS) $\RZ{V}=(\RZ{i}:i\in V)$ with $P_{\RZ{V}}$ being the joint probability mass function. We consider the problem of how to encode the sources in the DMMS $\RZ{V}$ so that the compressed data can be forwarded as a feasible flow in the digraph $G$ to the cluster header/sink $t$.
Note, in this problem, each node $i$ can generate and relay/forward information at the same time. Therefore, we have the constraints that are imposed by both the data compression of $\RZ{V}$ and the capacity function $c$ in the digraph $G$.

\section{Problem Formulation}
\label{sec:ProbForm}
For a flow $\flow$, define the \textit{boundary} $\boundary \colon 2^V \mapsto \Real$ by \cite[Section 1.2]{Fujishige2005}
$$ \boundary (X) = \sum_{(i,j) \in E \colon i \in X} \flow(i,j) - \sum_{(i,j) \in E \colon j \in X} \flow(i,j)$$
for all $X \subseteq V$. Here, $\sum_{(i,j) \in E \colon i \in X} \flow(i,j)$ and $\sum_{(i,j) \in E \colon j \in X} \flow(i,j)$ quantify the total incoming and outgoing information flow to and from the node set $X$, respectively. Note, $\sum_{(i,j) \in E \colon j \in X} \flow(i,j)$ is the amount of the compressed data flow from $V \setminus X$ and is supposed to be forwarded by $X$. Then, $\boundary(X)$ denotes the source coding rate that is assigned by the flow $\flow$ to encode the source $\RZ{X}$ and $\Boundary = (\boundary(\Set{i}) \colon i \in V)$ is the source coding vector designated by the flow $\flow$ to encode the DMMS $\RZ{V}$.

For $X \subseteq V$, let $H(X)$ be the amount of randomness in $\RZ{X}$ measured by Shannon entropy \cite{Cover2012ITBook}. Then, the maximum independent information amount that can be obtained by the source coding rate $\boundary(X)$ is upper bounded by $H(X)$, i.e., $\boundary(X) \leq H(X), \forall X \subseteq V$, and all flows $\flow$ in the digraph $G$ that result in a source coding rate vector $\Boundary$ at which we can source independent randomness from the DMMS $\RZ{V}$ is constrained by $\Boundary \in P(H,\leq)$, where
$$ P(H,\leq) = \Set{\Boundary \in \RealP^{|V|} \colon \boundary(X) \leq H(X), \forall X \subseteq V} $$
is the \textit{polyhedron} of $H$. Note, when we set the sum-rate $\boundary(V) = H(V)$, the constraints in $P(H,\leq)$ can be converted to $\boundary(X) \geq H(X | V \setminus X), \forall X \subseteq V$ so that $\Boundary \in P(H,\leq)$ is equivalent to the Slepian-Wolf constraints \cite{SW1973,Cover1975} for the lossless data compression of $\RZ{V}$.\footnote{For the data compression problem, the objective is to minimize the information redundancy when considering the Slepian-Wolf constraints $\boundary(X) \geq H(X | V \setminus X),\forall X \subseteq V$ and to minimize the information loss when considering the constraints $\boundary(X) \leq H(X),\forall X \subseteq V$ in the polyhedron $P(H,\leq)$. }

The objective is to find a feasible flow in the digraph $G$ such that we can source the maximum amount of independent information from $V$ to $t$:
\begin{equation} \label{eq:Prob}
        \begin{aligned}
        \max & \quad \boundary(V)  \\
        \text{s.t.} & \quad 0 \leq \flow(i,j) \leq c(i,j), \forall (i,j) \in E  \\
                    & \quad \Boundary \in P(H,\leq).
        \end{aligned}
\end{equation}

\section{Maximum Independent Flow Algorithm}
\label{sec:Solution}

The maximization in \eqref{eq:Prob} is called \textit{maximum independent flow (MIF)} problem and can be directly solved by a recursive algorithm \cite[Section 7]{FujishigeMaxIndFlow1978}. In this section, we adapt this MIF algorithm for solving problem~\eqref{eq:Prob} so that it can be implemented in a distributed manner. We explain that, when the MIF algorithm applies to \eqref{eq:Prob}, it actually repeatedly pushes the remaining randomness in $\RZ{V}$ over the digraph $G$ to the sink.

For a feasible flow $\flow$ in the digraph $G$ for the MIF problem~\eqref{eq:Prob}, i.e., $\flow$ satisfies the constraints in \eqref{eq:Prob}, and the resulting source coding rate vector $\Boundary$, the \textit{saturation capacity} to each dimension $i \in V$ is \cite[Section 2.2]{Fujishige2005}
$$ \hat{c}(\Boundary,i) = \max \Set{\alpha \colon \Boundary + \alpha \chi_i \in P(H,\leq)}, $$
where $\chi_i \in \Z^{|V|}$ is the \textit{characteristic vector} with the $i$th dimension being $1$ and all other dimensions being $0$.
The saturation capacity $\hat{c}(\Boundary,i)$ measures the remaining randomness in $\RZ{i}$ given the compressed data that has flowed to the sink $t$ via $\flow$. So, if $\hat{c}(\Boundary,i) = 0$, dimension $i$ is saturated, i.e., we can not source any more randomness from node $i$. For the saturated dimensions $i,j$, we have the \textit{exchange capacity} \cite[Section 2.2]{Fujishige2005}
$$ \hat{c}(\Boundary,i,j) = \max \Set{\alpha \colon \Boundary + \alpha (\chi_i -\chi_j) \in P(H,\leq)}.$$
Here, if $\hat{c}(\Boundary,i,j) > 0$, we can transfer at most $\hat{c}(\Boundary,i,j)$ source coding rates from node $j$ to node $i$. This is apparently due to the mutual dependence between $\RZ{i}$ and $\RZ{j}$: It makes no difference for either node to reveal the shared information. See Example~\ref{ex:NonInt}. Then, the \textit{dependence function}
$$ \dep(\Boundary,i) = \begin{cases} \Set{j \in V \colon \hat{c}(\Boundary,i,j) > 0 } & \hat{c}(\Boundary,i) = 0 \\ \emptyset & \hat{c}(\Boundary,i) = 0 \end{cases}.$$
determines all nodes that can exchange source coding rates with a saturated node $i$.

       \begin{algorithm} [t]
	       \label{algo:MIF}
	       \small
	       \SetAlgoLined
	       \SetKwInOut{Input}{input}\SetKwInOut{Output}{output}
	       \SetKwFor{For}{for}{do}{endfor}
            \SetKwRepeat{Repeat}{repeat}{until}
            \SetKwIF{If}{ElseIf}{Else}{if}{then}{else if}{else}{endif}
	       \BlankLine
           \Input{A flow that satisfies the constraints in \eqref{eq:Prob}, e.g., a zero flow $\flow(i,j) = 0, \forall (i,j) \in E$ in $G$. }
	       \Output{An optimal flow $\flow$ to problem~\eqref{eq:Prob}.}
	       \BlankLine
            \Repeat{$\boundary(V) = H(V)$}{
                \ForEach{$i \in V$}{$\hat{c}(
                    \Boundary,i) \leftarrow \max \Set{\alpha \colon \Boundary + \alpha \chi_i \in P(H,\leq)}$\;
                    \lIf{$\hat{c}(\Boundary,i) > 0$}{search a shortest path $\rho_i$ from $i$ to $t$ in $G_{\flow}$}
                    }
                \lIf{no $\rho_i$ is found}{terminate iteration and go to step~\ref{step:returnNonInt}} \label{step:TermNonInt}
                Let $\hat{\rho}$ be the $\rho_i$ with shortest length and smallest index $\hat{i}$\;
                $\beta \leftarrow \min\Set{ \hat{c}(\Boundary,\hat{i}), \min\Set{c(i,j) \colon (i,j) \in \hat{\rho}} }$\;
                \lForEach{$(i,j) \in \hat{\rho}$}{
                    $ \flow(i,j)\leftarrow \begin{cases} \flow(i,j) + \beta & (i,j) \in E_{\flow}^+ \\ \flow(i,j) - \beta & (i,j) \in E_{\flow}^- \end{cases}$
                }
            }
            return $\flow$\;\label{step:returnNonInt}
	   \caption{Maximum Independent Flow (MIF) Algorithm: A distributed implementation}
	   \end{algorithm}

The MIF algorithm is shown in Algorithm~\ref{algo:MIF}, where $G_{\flow} = ( V \cup \Set{t}, E_{\flow}^+ \cup E_{\flow}^- \cup D_{\flow},c_{\flow})$ is an auxiliary digraph with the edge sets and capacity function being
\begin{equation}
    \begin{aligned}
        & E_{\flow}^+ = \Set{(i,j) \colon (i,j) \in E, \flow(i,j) < c(i,j)}; \\
        & E_{\flow}^- = \Set{(j,i) \colon (i,j) \in E, \flow(i,j) > 0 }; \\
        & D_{\flow} = \Set{(i,j) \colon i \in \dep(\Boundary,j) \setminus \Set{j} }; \\
        & c_{\flow}(i,j) =  \begin{cases} c(i,j) - \flow(i,j) & (i,j) \in E_{\flow}^+ \\ \flow(j,i) & (i,j) \in E_{\flow}^- \\ \hat{c}(\Boundary, j,i) & (i,j) \in D_{\flow} \end{cases}.
    \end{aligned} \nonumber
\end{equation}
The edge sets $E_{\flow}^+$ and $E_{\flow}^-$ are due to the edge capacities in the digraph $G$: The flow $\flow$ remains feasible if we increase $\flow(i,j)$ by $c(i,j) - \flow(i,j)$ or reduce $\flow(i,j)$ by $\flow(i,j)$. The edge set $D_{\flow}$ is due to the nonzero exchange capacity $\hat{c}(\Boundary, j,i)$. So, $G_{\flow}$ characterizes all increments on flow $\flow$ and the exchanges of source coding rates between nodes such that the resulting flow remains feasible for problem~\eqref{eq:Prob}.

If, for some node $i$ such that $\hat{c}(\Boundary,i) > 0$, there exists a directed path $\rho_i$ in $G_{\flow}$ from $i$ to the sink $t$, we can push the remaining randomness in $\RZ{i}$ towards $t$ over path $\rho_i$ and the maximum flow increment is $\beta = \min\Set{ \hat{c}(\Boundary,i), \min\Set{c(i,j) \colon (i,j) \in \rho_i} }$ \cite[Theorem 2]{FujishigeMaxIndFlow1978}. Also, for all edges $(i,j)$ in the path $\rho_i$ such that $(i,j) \in D_{\flow}$, i.e., $i \in \dep(\Boundary,j)$, there are $\beta$ source coding rates transferred from $i$ to $j$. See Example~\ref{ex:NonInt}. A flow $\flow$ is the optimal solution to \eqref{eq:Prob} if there does not exist any directed path from any unsaturated node $i$ to $t$ \cite[Theorem 4]{FujishigeMaxIndFlow1978}. So, the MIF algorithm recursively push the remaining randomness in the source nodes via the increment of the flow and/or the exchange of the source coding rates until it reaches the optimal flow.\footnote{Steps 6 and 7 in Algorithm~\ref{algo:MIF} seek the lexicographically shortest path in $G_{\flow}$. It ensures the finiteness of the recursions in the MIF algorithm\cite[Theorem 4.11]{Fujishige2005}\cite{Lawler1982}.}

\begin{figure}[tbp]
	\centering
        \subfigure[$\Boundary = (0,0,0,0)$]{\scalebox{0.75}{\begin{tikzpicture}

\draw (0,0) circle (0.3);
\node at (0,0) {\Large $2$};

\draw (1.5,1) circle (0.3);
\node at (1.5,1) {\Large $3$};

\draw (1.5,-1) circle (0.3);
\node at (1.5,-1) {\Large $t$};

\draw (-1.5,1) circle (0.3);
\node at (-1.5,1) {\Large $1$};

\draw (-1.5,-1) circle (0.3);
\node at (-1.5,-1) {\Large $4$};

\draw [->] (-1.2,1) -- node [auto] {\scriptsize $0/2$} (1.2,1);
\draw [->] (1.5,0.7) -- node [auto] {\scriptsize $0/2$} (1.5,-0.7);
\draw [->] (-1.3,0.8) -- node [sloped,above] {\scriptsize $0/1$} (-0.25,0.15);
\draw [->] (-1.5,0.7) -- node [auto,left] {\scriptsize $0/3$} (-1.5,-0.7);
\draw [->] (-1.3,-0.8) -- node [sloped,above] {\scriptsize $0/1$} (-0.25,-0.15);
\draw [->] (0.25,-0.15) -- node [sloped,above] {\scriptsize $0/0.6$}(1.3,-0.8);
\draw [->] (1.2,-1) -- node [above] {\scriptsize $0/1$}(-1.2,-1);

\end{tikzpicture} }}\qquad
		\subfigure[$\Boundary = (0,0.6,0,0)$]{\scalebox{0.75}{\begin{tikzpicture}

\draw (0,0) circle (0.3);
\node at (0,0) {\Large $2$};

\draw (1.5,1) circle (0.3);
\node at (1.5,1) {\Large $3$};

\draw (1.5,-1) circle (0.3);
\node at (1.5,-1) {\Large $t$};

\draw (-1.5,1) circle (0.3);
\node at (-1.5,1) {\Large $1$};

\draw (-1.5,-1) circle (0.3);
\node at (-1.5,-1) {\Large $4$};

\draw [->] (-1.2,1) -- node [auto] {\scriptsize $0/2$} (1.2,1);
\draw [->] (1.5,0.7) -- node [auto] {\scriptsize $0/2$} (1.5,-0.7);
\draw [->] (-1.3,0.8) -- node [sloped,above] {\scriptsize $0/1$} (-0.25,0.15);
\draw [->] (-1.5,0.7) -- node [auto,left] {\scriptsize $0/3$} (-1.5,-0.7);
\draw [->] (-1.3,-0.8) -- node [sloped,above] {\scriptsize $0/1$} (-0.25,-0.15);
\draw [->] (0.25,-0.15) -- node [sloped,above] {\scriptsize $\color{red}{0.6}$$/0.6$}(1.3,-0.8);
\draw [->] (1.2,-1) -- node [above] {\scriptsize $0/1$}(-1.2,-1);

\end{tikzpicture} }}\\
        \subfigure[$\Boundary = (1,0.6,0,0)$]{\scalebox{0.75}{\begin{tikzpicture}

\draw (0,0) circle (0.3);
\node at (0,0) {\Large $2$};

\draw (1.5,1) circle (0.3);
\node at (1.5,1) {\Large $3$};

\draw (1.5,-1) circle (0.3);
\node at (1.5,-1) {\Large $t$};

\draw (-1.5,1) circle (0.3);
\node at (-1.5,1) {\Large $1$};

\draw (-1.5,-1) circle (0.3);
\node at (-1.5,-1) {\Large $4$};

\draw [->] (-1.2,1) -- node [auto] {\scriptsize $\color{red}{1}$$/2$} (1.2,1);
\draw [->] (1.5,0.7) -- node [auto] {\scriptsize $\color{red}{1}$$/2$} (1.5,-0.7);
\draw [->] (-1.3,0.8) -- node [sloped,above] {\scriptsize $0/1$} (-0.25,0.15);
\draw [->] (-1.5,0.7) -- node [auto,left] {\scriptsize $0/3$} (-1.5,-0.7);
\draw [->] (-1.3,-0.8) -- node [sloped,above] {\scriptsize $0/1$} (-0.25,-0.15);
\draw [->] (0.25,-0.15) -- node [sloped,above] {\scriptsize $0.6/0.6$}(1.3,-0.8);
\draw [->] (1.2,-1) -- node [above] {\scriptsize $0/1$}(-1.2,-1);

\end{tikzpicture} }} \qquad
        \subfigure[$\Boundary = (1,0.2,0.4,0.4)$]{\scalebox{0.75}{\begin{tikzpicture}

\draw (0,0) circle (0.3);
\node at (0,0) {\Large $2$};

\draw (1.5,1) circle (0.3);
\node at (1.5,1) {\Large $3$};

\draw (1.5,-1) circle (0.3);
\node at (1.5,-1) {\Large $t$};

\draw (-1.5,1) circle (0.3);
\node at (-1.5,1) {\Large $1$};

\draw (-1.5,-1) circle (0.3);
\node at (-1.5,-1) {\Large $4$};

\draw [->] (-1.2,1) -- node [auto] {\scriptsize $1/2$} (1.2,1);
\draw [->] (1.5,0.7) -- node [auto] {\scriptsize $\color{red}{1.4}$$/2$} (1.5,-0.7);
\draw [->] (-1.3,0.8) -- node [sloped,above] {\scriptsize $0/1$} (-0.25,0.15);
\draw [->] (-1.5,0.7) -- node [auto,left] {\scriptsize $0/3$} (-1.5,-0.7);
\draw [->] (-1.3,-0.8) -- node [sloped,above] {\scriptsize $\color{red}{0.4}$$/1$} (-0.25,-0.15);
\draw [->] (0.25,-0.15) -- node [sloped,above] {\scriptsize $0.6/0.6$}(1.3,-0.8);
\draw [->] (1.2,-1) -- node [above] {\scriptsize $0/1$}(-1.2,-1);

\end{tikzpicture} }}
	\caption{The updates of the flow $\flow$, presented as $f(i,j)/c(i,j)$ on each edge, and the resulting source coding vector $\Boundary$ at each iteration of the MIF algorithm when it is applied to digraph in Fig.~\ref{fig:Digraph} where the sensors in $V = \Set{1,\dotsc,4}$ observes a DMMS $\RZ{V}$ in Example~\ref{ex:NonInt}. The flows in red are the updated ones from the last iteration.
}
	\label{fig:ExNonInt}
\end{figure}
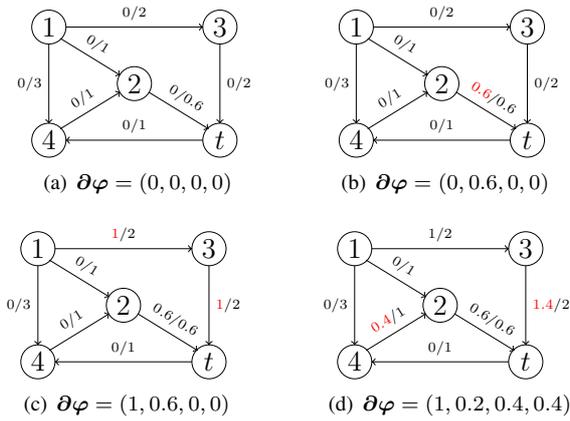

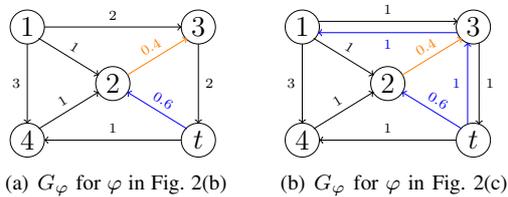
\begin{figure}[tbp]
	\centering
        \subfigure[$G_{\flow}$ for $\flow$ in Fig.~\ref{fig:ExNonInt}(b)]{\scalebox{0.75}{\begin{tikzpicture}

\draw (0,0) circle (0.3);
\node at (0,0) {\Large $2$};

\draw (1.5,1) circle (0.3);
\node at (1.5,1) {\Large $3$};

\draw (1.5,-1) circle (0.3);
\node at (1.5,-1) {\Large $t$};

\draw (-1.5,1) circle (0.3);
\node at (-1.5,1) {\Large $1$};

\draw (-1.5,-1) circle (0.3);
\node at (-1.5,-1) {\Large $4$};

\draw [->] (-1.2,1) -- node [auto] {\scriptsize $2$} (1.2,1);
\draw [->] (1.5,0.7) -- node [auto] {\scriptsize $2$} (1.5,-0.7);
\draw [->] (-1.3,0.8) -- node [sloped,above] {\scriptsize $1$} (-0.25,0.15);
\draw [->] (-1.5,0.7) -- node [auto,left] {\scriptsize $3$} (-1.5,-0.7);
\draw [->] (-1.3,-0.8) -- node [sloped,above] {\scriptsize $1$} (-0.25,-0.15);
\draw [->,blue] (1.3,-0.8) -- node [sloped,above] {\scriptsize $0.6$}(0.25,-0.15);
\draw [->] (1.2,-1) -- node [above] {\scriptsize $1$}(-1.2,-1);

\draw [->,orange] (0.25,0.15) -- node [sloped,above] {\scriptsize $0.4$}(1.3,0.8);

\end{tikzpicture} }}\qquad
		\subfigure[$G_{\flow}$ for $\flow$ in Fig.~\ref{fig:ExNonInt}(c)]{\scalebox{0.75}{\begin{tikzpicture}

\draw (0,0) circle (0.3);
\node at (0,0) {\Large $2$};

\draw (1.5,1) circle (0.3);
\node at (1.5,1) {\Large $3$};

\draw (1.5,-1) circle (0.3);
\node at (1.5,-1) {\Large $t$};

\draw (-1.5,1) circle (0.3);
\node at (-1.5,1) {\Large $1$};

\draw (-1.5,-1) circle (0.3);
\node at (-1.5,-1) {\Large $4$};

\draw [->] (-1.22,1.1) -- node [auto] {\scriptsize $1$} (1.22,1.1);
\draw [->,blue] (1.22,0.9) -- node [auto] {\scriptsize $1$} (-1.22,0.9);

\draw [->] (1.6,0.72) -- node [auto] {\scriptsize $1$} (1.6,-0.72);
\draw [->,blue] (1.4,-0.72) -- node [auto] {\scriptsize $1$} (1.4,0.72);

\draw [->] (-1.3,0.8) -- node [sloped,above] {\scriptsize $1$} (-0.25,0.15);
\draw [->] (-1.5,0.7) -- node [auto,left] {\scriptsize $3$} (-1.5,-0.7);
\draw [->] (-1.3,-0.8) -- node [sloped,above] {\scriptsize $1$} (-0.25,-0.15);
\draw [->,blue] (1.3,-0.8) -- node [sloped,above] {\scriptsize $0.6$}(0.25,-0.15);
\draw [->] (1.2,-1) -- node [above] {\scriptsize $1$}(-1.2,-1);

\draw [->,orange] (0.25,0.15) -- node [sloped,above] {\scriptsize $0.4$}(1.3,0.8);

\end{tikzpicture} }}
	\caption{The auxiliary digraph $G_{\flow}$ at the $2$nd and $3$rd iterations of the MIF algorithm in Example~\ref{ex:NonInt}, where the edge $(2,3) \in D_{\flow}$ is due to the nonzero exchange capacity $\hat{c}(\Boundary,3,2) = 0.4$.
Note, $\hat{c}(\Boundary,3,2) = 0.4$ is resulted from the mutual dependence between $\RZ{2}$ and $\RZ{3}$: $I(\Set{2} \wedge \Set{3}) = 0.4$.  }
	\label{fig:AuxGEx}
\end{figure}

\begin{example} \label{ex:NonInt}
For the digraph in Fig.~\ref{fig:Digraph} with $V = \Set{1,\dotsc,4}$, let dimensions in the DMMS $\RZ{V}$ be
\begin{equation}
    \begin{aligned}
        \RZ{1} &= (\RW{a},\RW{b}), \quad  & \RZ{2} = (\RW{b},\RW{c}),   \nonumber\\
        \RZ{3} &= (\RW{c}), \quad  & \RZ{4} = (\RW{b},\RW{d})
    \end{aligned}
\end{equation}
where, for all $m \in \Set{1,\dotsc,4}$, $\RW{m}$ is an independent random bit with $H(\RW{a}) = 1$, $H(\RW{b}) = 0.2$ and $H(\RW{c}) = H(\RW{d}) = 0.4$. We start the MIF algorithm with zero flow $\flow$, $\flow(i,j) = 0, \forall (i,j) \in E$ as shown in Fig.~\ref{fig:ExNonInt}(a). The source coding rate vector determined by the boundary is $\Boundary = (0,0,0,0)$.

At the $1$st iteration, since we have not pushed any information to the sink $t$, the saturation capacity is $\hat{c}(\Boundary,i) = H(\Set{i}) > 0$ for all $i \in V$, i.e., we have nonzero remaining randomness at all source nodes. Also, $G_{\flow} = G$ and $\hat{\rho} = \rho_2 = (2,t)$ is the shortest source-to-sink path over all $i \in V$ and $\beta = \min \Set{ \hat{c}(\Boundary,2), c(2,t) } = 0.6$. We increase $f(2,t)$ by $0.6$ which results in a flow in Fig.~\ref{fig:ExNonInt}(b). The corresponding source coding rate vector is $\Boundary = (0,0.6,0,0)$.

At the $2$nd iteration, we have $\hat{c}(\Boundary,1) = 1$, $\hat{c}(\Boundary,4) = 0.4$ and $\hat{c}(\Boundary,2) = \hat{c}(\Boundary,3) = 0$. The auxiliary digraph $G_{\flow}$ is shown in Fig.~\ref{fig:AuxGEx}(a). We have $\hat{\rho} = \rho_1 = (1,3) \rightarrow (3,t)$ being the shortest path from unsaturated source set $\Set{1,4}$ to $t$ and $\beta = 1$. We increase $\flow(1,3)$ and $\flow(3,t)$ by $1$, i.e., push $1$ bit of randomness from node $1$ to $t$, and results in a flow in Fig.~\ref{fig:ExNonInt}(c).

At the $3$rd iteration, we have node $4$ being the only unsaturated source node with the remaining randomness $\hat{c}(\Boundary,4) = 0.4$ and $G_{\flow}$ in Fig.~\ref{fig:AuxGEx}(b). Note, the edge $(2,3) \in D_{\flow}$ with the exchange capacity $\hat{c}(\Boundary,3,2) = 0.4$ is because of the mutual information $I(\Set{2} \wedge \Set{3}) = 0.4$: There are $0.4$ bit of shared information that can be transmitted by either $2$ or $3$ and, therefore, node $2$ can transfer at most $0.4$ source coding rates to node $3$. In $G_{\flow}$, $\hat{\rho} = \rho_4 = (4,2) \rightarrow (2,3) \rightarrow (3,t)$ is the only, and also shortest, path from $4$ to $t$ and $\beta = 0.4$. Since the edge $(2,3) \in \hat{\rho}$, when we push $\beta = 0.4$ over $\hat{\rho}$, what happens in the original graph $G$ is that we reduce $\boundary(\Set{2})$ by $0.4$ and increase $\boundary(\Set{3})$ by $0.4$, which results in a flow in Fig.~\ref{fig:ExNonInt}(c) with the source coding rate vector being $\Boundary = (1,0.2,0.4,0.4)$. Now, we have $\boundary(V) = 2 = H(V)$ and the MIF algorithm terminates with the flow $\flow$ updated to the optimum.\footnote{One can verify that the source coding rate vector $\Boundary = (1,0.2,0.4,0.4)$ also satisfies the Slepian-Wolf constraints \cite{SW1973,Cover1975}.}
\end{example}

\subsection{Complexity and Distributed Implementation}
\label{sec:ComplexNonInt}

In the MIF algorithm, the saturation and exchange capacities, $\hat{c}(\Boundary, i)$ and $\hat{c}(\Boundary, i, j), \forall i,j \in V$, can be determined by set function minimization problems \cite[Section 2.2]{Fujishige2005}
\begin{equation}
    \begin{aligned}
        & \max \Set{\alpha \colon \Boundary + \alpha \chi_i \in P(H,\leq)} \\
        & \qquad\quad\quad = \min\Set{ H(X) - \boundary(X) \colon X \subseteq V, i \in X}; \\
        & \max \Set{\alpha \colon \Boundary + \alpha (\chi_i -\chi_j) \in P(H,\leq)} \\
        & \qquad\quad\quad = \min\Set{H(X) - \boundary(X) \colon X \subseteq V, i \in X, j \notin X},
    \end{aligned}\nonumber
\end{equation}
where the two minimizations can be solved by the submodular function minimization (SFM) algorithms \cite[Chapter VI]{Fujishige2005} due to the submodularity of the entropy function $H$ \cite{FujishigePolyEntropy}. Since we need to obtain $\hat{c}(\Boundary, i, j)$ for each pair $(i,j)$ when $\hat{c}(\Boundary, i) \neq 0$, the complexity in each iteration of the MIF algorithm is upper bounded by $O(|V|^2 \cdot \SFM(|V|))$.\footnote{$O(\SFM(|V|))$ denotes the complexity of solving problem $\min \Set{H(X) - \boundary(X) \colon X \subseteq V}$ and ranges from $O(|V|^5)$ to $O(|V|^8)$ \cite[Chapter VI]{Fujishige2005}. Note, we neglect the complexity of the shortest path algorithm since it is much less complex than solving the SFM problem, e.g., the Dijkstra's algorithm \cite{Dijkstra1959} searches a shortest path in $O(|V|^2)$ time. }
Also, the total number of iterations in the MIF algorithm is no greater than $|V|^3$ \cite[Theorem 4.11]{Fujishige2005}. The MIF algorithm completes in $O(|V|^5 \cdot \SFM(|V|))$ time.

The MIF algorithm in Algorithm~\ref{algo:MIF} implies a decentralized computation method: Each node $i$ obtains its own capacities $\hat{c}(\Boundary, i)$ and $\hat{c}(\Boundary, i, j)$; Most of the shortest path algorithms, e.g., \cite{Dijkstra1959}, allows distributed implementation where each node only needs to know the connection in the neighborhood; The nodes can negotiate with each other to determine $\hat{\rho}$. Then, the complexity at each node is $O(|V|^4 \cdot \SFM(|V|))$.

\section{Integral Maximum Independent Flow Algorithm}
It can be seen from Section~\ref{sec:ComplexNonInt} that calculating the saturation and exchange capacities consumes most of the computation capacity in the MIF algorithm. Thus, it is worth discussing how to simplify or avoid the computation of $\hat{c}(\Boundary, i)$ and $\hat{c}(\Boundary, i, j)$. We show in this section that this is possible if the capacities $c(i,j)$ in the digraph $G$ are integral and the entropy $H$ of the DMMS $\RZ{V}$ is integer-valued. In fact, the integrity of $c$ and $H$ reduces \eqref{eq:Prob} to a network coding problem in a network.

For the digraph $G = \Set{ V \cup \Set{t}, E, c}$ with $c \colon E \mapsto \Z_{++}$ and the DMMS $\RZ{V}$ with $H \colon 2^V \mapsto \Z_+$, there exists a flow $\flow \colon E \mapsto \Z_{+}$ that optimizes problem~\eqref{eq:Prob} \cite[Theorem 5]{FujishigeMaxIndFlow1978}. Inspired by the idea of the MIF algorithm, we can obtain this optimal integral flow by starting with the zero flow and keeping pushing unit remaining randomness until we cannot do so any more. By doing so, we can reduce the auxiliary digraph $G_{\flow}$ to an uncapacitated one $G_{\flow}^I = ( V \cup \Set{t}, E_{\flow}^+ \cup E_{\flow}^- \cup D_{\flow})$. Then, we have the integral maximum independent flow (IMIF) algorithm in Algorithm~\ref{algo:IMIF}.

       \begin{algorithm} [t]
	       \label{algo:IMIF}
	       \small
	       \SetAlgoLined
	       \SetKwInOut{Input}{input}\SetKwInOut{Output}{output}
	       \SetKwFor{For}{for}{do}{endfor}
            \SetKwRepeat{Repeat}{repeat}{until}
            \SetKwIF{If}{ElseIf}{Else}{if}{then}{else if}{else}{endif}
	       \BlankLine
           \Input{A zero flow $\flow(i,j) = 0, \forall (i,j) \in E$ in $G$. }
	       \Output{An integral optimal flow $\flow$ to problem~\eqref{eq:Prob}.}
	       \BlankLine
            \Repeat{$\boundary(V) = H(V)$}{
                \lForEach{$i \in V$ such that $\hat{c}(\Boundary,i) > 0$}{ search a shortest path $\rho_i$ from $i$ to $t$ in $G_{\flow}^I$}
                \lIf{no $\rho_i$ is found}{terminate iteration and go to step~\ref{step:returnInt}} \label{step:TermInt}
                Let $\hat{\rho}$ be the $\rho_i$ with shortest length and smallest index $i$\;
                \lForEach{$(i,j) \in \hat{\rho}$}{
                    $ \flow(i,j)\leftarrow \begin{cases} \flow(i,j) + 1 & (i,j) \in E_{\flow}^+ \\ \flow(i,j) - 1 & (i,j) \in E_{\flow}^- \end{cases}$
                }
            }
            return $\flow$\label{step:returnInt}\;
	   \caption{Integral Maximum Independent Flow (IMIF) Algorithm}
	   \end{algorithm}

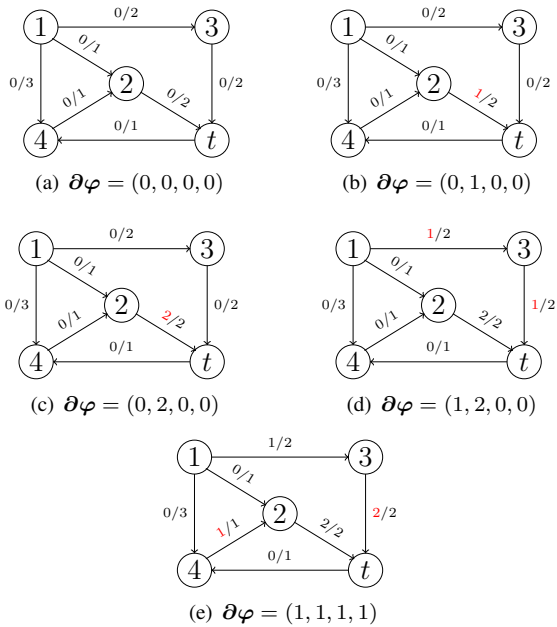
\begin{figure}[tbp]
	\centering
        \subfigure[$\Boundary = (0,0,0,0)$]{\scalebox{0.75}{\begin{tikzpicture}

\draw (0,0) circle (0.3);
\node at (0,0) {\Large $2$};

\draw (1.5,1) circle (0.3);
\node at (1.5,1) {\Large $3$};

\draw (1.5,-1) circle (0.3);
\node at (1.5,-1) {\Large $t$};

\draw (-1.5,1) circle (0.3);
\node at (-1.5,1) {\Large $1$};

\draw (-1.5,-1) circle (0.3);
\node at (-1.5,-1) {\Large $4$};

\draw [->] (-1.2,1) -- node [auto] {\scriptsize $0/2$} (1.2,1);
\draw [->] (1.5,0.7) -- node [auto] {\scriptsize $0/2$} (1.5,-0.7);
\draw [->] (-1.3,0.8) -- node [sloped,above] {\scriptsize $0/1$} (-0.25,0.15);
\draw [->] (-1.5,0.7) -- node [auto,left] {\scriptsize $0/3$} (-1.5,-0.7);
\draw [->] (-1.3,-0.8) -- node [sloped,above] {\scriptsize $0/1$} (-0.25,-0.15);
\draw [->] (0.25,-0.15) -- node [sloped,above] {\scriptsize $0/2$}(1.3,-0.8);
\draw [->] (1.2,-1) -- node [above] {\scriptsize $0/1$}(-1.2,-1);

\end{tikzpicture} }}\qquad
		\subfigure[$\Boundary = (0,1,0,0)$]{\scalebox{0.75}{\begin{tikzpicture}

\draw (0,0) circle (0.3);
\node at (0,0) {\Large $2$};

\draw (1.5,1) circle (0.3);
\node at (1.5,1) {\Large $3$};

\draw (1.5,-1) circle (0.3);
\node at (1.5,-1) {\Large $t$};

\draw (-1.5,1) circle (0.3);
\node at (-1.5,1) {\Large $1$};

\draw (-1.5,-1) circle (0.3);
\node at (-1.5,-1) {\Large $4$};

\draw [->] (-1.2,1) -- node [auto] {\scriptsize $0/2$} (1.2,1);
\draw [->] (1.5,0.7) -- node [auto] {\scriptsize $0/2$} (1.5,-0.7);
\draw [->] (-1.3,0.8) -- node [sloped,above] {\scriptsize $0/1$} (-0.25,0.15);
\draw [->] (-1.5,0.7) -- node [auto,left] {\scriptsize $0/3$} (-1.5,-0.7);
\draw [->] (-1.3,-0.8) -- node [sloped,above] {\scriptsize $0/1$} (-0.25,-0.15);
\draw [->] (0.25,-0.15) -- node [sloped,above] {\scriptsize $\color{red}{1}$$/2$}(1.3,-0.8);
\draw [->] (1.2,-1) -- node [above] {\scriptsize $0/1$}(-1.2,-1);

\end{tikzpicture} }}\\
        \subfigure[$\Boundary = (0,2,0,0)$]{\scalebox{0.75}{\begin{tikzpicture}

\draw (0,0) circle (0.3);
\node at (0,0) {\Large $2$};

\draw (1.5,1) circle (0.3);
\node at (1.5,1) {\Large $3$};

\draw (1.5,-1) circle (0.3);
\node at (1.5,-1) {\Large $t$};

\draw (-1.5,1) circle (0.3);
\node at (-1.5,1) {\Large $1$};

\draw (-1.5,-1) circle (0.3);
\node at (-1.5,-1) {\Large $4$};

\draw [->] (-1.2,1) -- node [auto] {\scriptsize $0/2$} (1.2,1);
\draw [->] (1.5,0.7) -- node [auto] {\scriptsize $0/2$} (1.5,-0.7);
\draw [->] (-1.3,0.8) -- node [sloped,above] {\scriptsize $0/1$} (-0.25,0.15);
\draw [->] (-1.5,0.7) -- node [auto,left] {\scriptsize $0/3$} (-1.5,-0.7);
\draw [->] (-1.3,-0.8) -- node [sloped,above] {\scriptsize $0/1$} (-0.25,-0.15);
\draw [->] (0.25,-0.15) -- node [sloped,above] {\scriptsize $\color{red}{2}$$/2$}(1.3,-0.8);
\draw [->] (1.2,-1) -- node [above] {\scriptsize $0/1$}(-1.2,-1);

\end{tikzpicture} }} \qquad
        \subfigure[$\Boundary = (1,2,0,0)$]{\scalebox{0.75}{\begin{tikzpicture}

\draw (0,0) circle (0.3);
\node at (0,0) {\Large $2$};

\draw (1.5,1) circle (0.3);
\node at (1.5,1) {\Large $3$};

\draw (1.5,-1) circle (0.3);
\node at (1.5,-1) {\Large $t$};

\draw (-1.5,1) circle (0.3);
\node at (-1.5,1) {\Large $1$};

\draw (-1.5,-1) circle (0.3);
\node at (-1.5,-1) {\Large $4$};

\draw [->] (-1.2,1) -- node [auto] {\scriptsize $\color{red}{1}$$/2$} (1.2,1);
\draw [->] (1.5,0.7) -- node [auto] {\scriptsize $\color{red}{1}$$/2$} (1.5,-0.7);
\draw [->] (-1.3,0.8) -- node [sloped,above] {\scriptsize $0/1$} (-0.25,0.15);
\draw [->] (-1.5,0.7) -- node [auto,left] {\scriptsize $0/3$} (-1.5,-0.7);
\draw [->] (-1.3,-0.8) -- node [sloped,above] {\scriptsize $0/1$} (-0.25,-0.15);
\draw [->] (0.25,-0.15) -- node [sloped,above] {\scriptsize $2/2$}(1.3,-0.8);
\draw [->] (1.2,-1) -- node [above] {\scriptsize $0/1$}(-1.2,-1);

\end{tikzpicture} }}
        \subfigure[$\Boundary = (1,1,1,1)$]{\scalebox{0.75}{\begin{tikzpicture}

\draw (0,0) circle (0.3);
\node at (0,0) {\Large $2$};

\draw (1.5,1) circle (0.3);
\node at (1.5,1) {\Large $3$};

\draw (1.5,-1) circle (0.3);
\node at (1.5,-1) {\Large $t$};

\draw (-1.5,1) circle (0.3);
\node at (-1.5,1) {\Large $1$};

\draw (-1.5,-1) circle (0.3);
\node at (-1.5,-1) {\Large $4$};

\draw [->] (-1.2,1) -- node [auto] {\scriptsize $1/2$} (1.2,1);
\draw [->] (1.5,0.7) -- node [auto] {\scriptsize $\color{red}{2}$$/2$} (1.5,-0.7);
\draw [->] (-1.3,0.8) -- node [sloped,above] {\scriptsize $0/1$} (-0.25,0.15);
\draw [->] (-1.5,0.7) -- node [auto,left] {\scriptsize $0/3$} (-1.5,-0.7);
\draw [->] (-1.3,-0.8) -- node [sloped,above] {\scriptsize $\color{red}{1}$$/1$} (-0.25,-0.15);
\draw [->] (0.25,-0.15) -- node [sloped,above] {\scriptsize $2/2$}(1.3,-0.8);
\draw [->] (1.2,-1) -- node [above] {\scriptsize $0/1$}(-1.2,-1);

\end{tikzpicture} }}
	\caption{The updates of the flow $\flow$, presented as $f(i,j)/c(i,j)$ on each edge, and the resulting source coding vector $\Boundary$ at each iteration of the IMIF algorithm in Example~\ref{ex:Int}.}
	\label{fig:ExInt}
\end{figure}

\begin{example} \label{ex:Int}
For the digraph in Fig.~\ref{fig:Digraph}, we replace the capacity $c(2,t)$ by $2$ and assume that all $\RW{m}$ observed in the DMMS $\RZ{V}$ are independent uniformly random bit, i.e., $H(\RW{m}) = 1,\forall m \in \Set{a,b,c,d}$. We start the IMIF algorithm with zero flow $\flow$. The flow updates are shown in Fig.~\ref{fig:ExInt}, where we can see that the IMIF recursively pushes a unit randomness to the sink $t$ until an optimal integral flow in Fig.~\ref{fig:ExInt}(e) is fetched.
\end{example}

\subsection{Complexity and Distributed Implementation}
The saturation capacity $\hat{c}(\Boundary,i)$ and the edge set $D_{\flow}$ can be both determined by solving the SFM problem
\begin{equation} \label{eq:IMIF}
    \min\Set{ H(X) - \boundary(X) \colon X \subseteq V, i \in X}:
\end{equation}
$\hat{c}(\Boundary,i)$ is the maximum of \eqref{eq:IMIF}; $\dep(\Boundary,i)$ is the minimal minimizer of \eqref{eq:IMIF}, based on which, $D_{\flow}$ can be constructed. There are at most $H(V)$ iterations in Algorithm~\ref{algo:IMIF}. Therefore, the IMIF algorithm completes in $O(H(V) \cdot |V| \cdot \SFM(|V|))$ time. It can also be implemented in a distributed manner so that the complexity at each node is $O(H(V) \cdot \SFM(|V|))$.

\section{Submodular Intersection Problem}
\label{sec:SubMInterSect}

It can be seen that we cannot always direct the total information $H(V)$ of the DMMS to the sink. When the iteration terminates at step~\ref{step:TermNonInt} in the MIF algorithm, or step~\ref{step:TermInt} in the IMIF algorithm, it means that we still have remaining randomness in the source nodes that is unable to be pushed to the sink $t$. For example, for the digraph $G$ in Fig.~\ref{fig:Digraph}, if $c(3,t) = 1$, the maximum information amount that we can source from $V$ to $t$ is only $1.6$.

Let $\Cut(X) = \sum_{(i,j) \in E \colon i \in X} c(i,j), \forall X \in V$ be the \textit{cut function} of the digraph $G$ \cite[Section 1.2]{Fujishige2005}. Define the \textit{characteristic function} \cite[Section 3]{Megiddo1974}
$$ f(X) = \min \Set{ \Cut(Y) \colon  X \subseteq Y \subseteq V }, \quad \forall X \subseteq V, $$
which can be considered as the min-cut between the super source node $X$ and the sink $t$. It is shown in \cite[Lemmas 4.1 and 3.2]{Megiddo1974} that the boundary of any feasible flow $\flow$ in $G$ is upper bounded by $\boundary(X) \leq f(X), \forall X \subseteq V$, i.e.,
$ \Boundary \in P(f,\leq)$, and $f$ is submodular. For instance, in Example~\ref{ex:NonInt}, one can verify that $\Boundary = (1,0.2,0.4,0.4)$ determined by the MIF algorithm also belongs to the polyhedron $P(f,\leq)$, where $f$ is the characteristic function that is determined by the capacities in the digraph in Fig.~\ref{fig:Digraph}.

So, the problem~\eqref{eq:Prob} is equivalent to
\begin{equation} \label{eq:SubMIntersect}
    \max \Set{\boundary(V) \colon \Boundary \in P(H,\leq) \cap P(f,\leq)}.
\end{equation}
If the maximum of \eqref{eq:SubMIntersect} is strictly less than $H(V)$, e.g., when $f(V) = \Cut(V) < H(V)$, then it is not possible to source all the information in $\RZ{V}$ to $t$. Therefore, it is worth discussing how to characterise the maximum of \eqref{eq:SubMIntersect} (without running the MIF algorithm), which is useful when we want to select the cluster header that can collect the most of sensing data in the cluster header.

In fact, problem~\eqref{eq:SubMIntersect} maximizes the independent flow in the intersection of polyhedra $P(H,\leq)$ and $P(f,\leq)$, where both $H$ and $f$ are submdodular functions. This is called the \textit{submodular intersection problem} and there exist results based on this problem that can be utilized to further improve the efficiency of solving the MIF problem~\eqref{eq:Prob}.

\section{Conclusion}
We studied the problem of how to source maximum randomness from multiple sources to a sink node as a feasible flow in a digraph. It describes the data compression problem in a sensor network/cluster. We adapted the MIF algorithm in a distributed manner to solve this problem and explained that the MIF algorithm recursively pushes the remaining randomness in the sources to the sink or cluster header until it cannot do so any more. We also showed that an integral optimal solution is less complex to determine and provided a novel IMIF algorithm to do so. We pointed out that the nature of the data compression problem in a sensor network is to maximize the flow in the intersection of two submodular polyhedra.

Finally, the study also directly leads to several directions for future work. By assigning each edge a weight that denotes the wireless link quality, it is of interest to determine a flow that minimize the sum-weight among the solutions to problem~\eqref{eq:Prob}. On the other hand, as the source coding solution that satisfies the Slepian-Wolf constraints is not unique, it is worth discussing how to attain the fairness in the solution set of \eqref{eq:Prob}. Also, as pointed out in Section~\ref{sec:SubMInterSect}, one can address how to utilize the existing submodular intersection techniques, e.g. \cite{Fujishige1992Intersect}, to enhance the efficiency of solving problem~\eqref{eq:Prob}.

\bibliographystyle{IEEEtran}
\bibliography{DCGraphBIB}

\end{document}